\documentclass[12pt,a4paper]{article}
\usepackage{dfttrob}
\usepackage[numbers,sort&compress]{natbib}
\usepackage{latexuseful2e}
\usepackage{amsmath}
\usepackage{graphicx}
\dfttnum{DFTT 25/2002\\August 2002}

\begin{document}

\title{Questions and Remarks About Clans in Multiparticle Dynamics}
\author{A. Giovannini and R. Ugoccioni\\
 \it Dipartimento di Fisica Teorica and I.N.F.N - Sezione di Torino\\
 \it Via P. Giuria 1, 10125 Torino, Italy}
\maketitle

\begin{abstract}
The fact that several important effects in multiparticle 
dynamics, on which QCD has
not yet satisfactory predictions, have been interpreted in terms of
the validity of negative binomial (Pascal) regularity 
and related clan properties at the level of
simpler substructures, raises intriguing questions on clan properties
in all classes of collisions, the main one being whether clans are
observable objects or merely a mathematical concept.
We approach this problem by studying clan masses and rapidity
distributions in each substructure for \ee\ annihilation and $hh$
collisions, and find that such properties can indeed
characterise the different components. These results support the idea
that clans could be observable, a challenging problem for future experiments.
\end{abstract}

\section{Introduction}
As well known, the concept of \textit{clan}
has been introduced in the eighties \cite{AGLVH:1} in
order to interpret the occurrence of the approximate NB (Pascal)
regularity in final charged particles multiplicity distributions
(MD's) of the full sample of events both in full phase space and in
restricted rapidity intervals in all classes of collisions in the GeV
region. 

Clans were defined as group of particles of common ancestor with at
least one particle per clan; by
assumption no correlations exist among clans.
Accordingly, the production process was understood as a two-step
process: to independently produced clans in the first step (they are
Poissonianly distributed), it follows a second step in which each clan
decays into final particles with a logarithmic distribution.
The average number of clans, $\Nbar$, in rapidity intervals at fixed
c.m.\ energy  and at various c.m.\ energies, as well as the average
number of particles per clan, $\nc$, characterised fully the
multiplicity distribution properties in various classes of high energy
collisions.
These two parameters are linked to standard NB (Pascal) distribution
parameters $\nbar$ (the average charged multiplicity) and $k$ ($1/k =
D^2/\nbar^2 - 1/\nbar$, $D$ being the dispersion of the MD) by two
non-trivial relations
\begin{equation}
	\Nbar = k\ln\left(1+\frac{\nbar}{k}\right) \qquad\text{and}\qquad
			\nc = \frac{\nbar}{\Nbar} .
\end{equation}

We learned \cite{Schmitz}, 
within the just recalled elementary interpretation, that
$\Nbar$ in \ee\ annihilation is larger at the same c.m.\ energy than
% at the same c.m. energy? e+e- 14-44 GeV, ppbar 31-62 GeV
in hadron-hadron collisions, and that $\nc$ is much smaller in the
former than in the latter case. In deep inelastic scattering (DIS),
the situation turned out to be intermediate between the previous two:
$\Nbar$ behaves as in hadron-hadron collisions but $\nc$ as in \ee\
annihilation. 
In addition, in all classes of collisions, clans are larger (they
contain more particles) in central rapidity intervals than in
peripheral ones.

These remarkable properties of clans were obtained in a quite simple
framework and suggested an interesting clan picture at parton level by
using generalised local parton-hadron duality (GLPHD) \cite{AGLVH:2}.
It was found that partonic clans behave as bremsstrahlung gluon jets
originated by the initial quark (the dominant vertex in \ee\
annihilation is $q\to q+g$) and are generated very probably at quite
low virtualities (this consideration explains the high number of clans
and the relatively low population of partons [particles] per clan in
this case).
This interpretation should be compared with what happens very
reasonably in the same picture 
in hadron-hadron collisions, where bremsstrahlung gluon
jets are thought to be
generated at quite high virtualities and to have a lot of
virtuality space for generating larger partonic cascades (the process
is dominated here by the gluon self-interaction vertex and stronger
colour exchanges).

More accurate analyses of final charged particles MD's 
at higher energies (at LEP \cite{DEL:1+DEL:2}, UA5 \cite{UA5:3}) revealed
violations of the regularity. A shoulder was seen in the MD's both in
full phase space and in rapidity intervals in \ee\ annihilation and
$hh$ collisions,  which were not described by a single
NB(Pascal)MD. 
The death of the regularity was celebrated as an expected and sound
fact. Experimental complexity was winning over theoretical simplicity.

A different school of thought pointed out that the NB (Pascal)
regularity was not dead in multiparticle dynamics, but simply that it
was working at a more fundamental level of investigation, i.e., at the
level of different substructures characterising various classes of
collisions \cite{hqlett:2}.
The violation of NB (Pascal) regularity in the full sample of events
in a high energy collision was considered as the indication of the
existence of substructures  (or different classes of events),
and the suggestion was to explore the validity of the
regularity in these substructures.
It was shown \cite{DEL:single,frascati97:AG}
that the regularity was surviving in
the separate 2- and 3-jet samples of events in \ee\ annihilation and,
presumably, in soft and semi-hard components in $hh$ collisions.

It turns out in fact that the weighted superposition of the mentioned
substructures, each described by a NB(Pascal) MD, reproduces
approximately three observed behaviours in final
charged particles MD's in both classes of collisions: the first one is
the mentioned shoulder effect in $P_n$ vs $n$ ($P_n$ is the MD)
\cite{DEL:single,Fug}; the second one
are $n$-oscillations in $H_n = K_n/F_n$ vs $n$ \cite{hqlett:2,L3:mangeol}
($F_n$ are factorial moments, $K_n$ cumulant moments), and the third
one is the general behaviour of the forward-backward multiplicity
correlation strength \cite{OPAL:FB,RU:FB}.

The fact that three important effects on which QCD has not
yet satisfactory predictions (the only claim is the onset of the hard
gluon vertex) and which can be interpreted in terms of the same cause,
i.e., the validity of the NB (Pascal) regularity at a more elementary
level of investigation than initially thought, raises intriguing
questions on clan properties in all classes of collisions,
the main one being: are clans observable, or is clan concept a
purely statistical one like cluster expansion in statistical
mechanics? 
In order to approach the problem, we decided to proceed by asking
ourselves the following preliminary questions.
\begin{itemize}

\item  Are clans massive objects?

\item If clans are massive, are clan masses different 
in different classes of events (or substructures)
in a given collision?

\item If clans are massive, what about clan masses in 
different classes of collisions?
\end{itemize}

\section{First question.}

A quantitative  answer to the first question has been given by A.Bialas
and  A. Szczerba \cite{Bialas:clanmass};
it was stimulated by the observed qualitative properties of clan structure 
analysis  when applied to  multiplicity distributions in rapidity intervals 
in hadron hadron collisions for the full sample of events collected by UA5 
Collaboration.
Here as discussed previously, the charged particle 
multiplicity distributions in rapidity intervals are 
of course of NB (Pascal) type,
$P_n^{\text{NB}}   (k,\nbar)$, with $k$ and $\nbar$
increasing with rapidity and with
particles generated by each independently produced clan 
according to a logarithmic distribution,
$P_n^{\text{L}} (\beta)$, with $\beta = \nbar/(\nbar+ k)$.

Two assumptions were at the basis of the mentioned generalisation
of standard  clan structure analysis to rapidity intervals:
they concern the distributions of clans in rapidity and the angular
distribution in clan decay respectively.

As previously mentioned, clans are Poissonianly distributed and independently
emitted in bremsstrahlung-like fashion. Using energy and
(longitudinal) momentum conservation, the single-clan 
(pseudo)-rapidity density has been
written \cite{Stodolski+DeGroot+BialasHayot+BialasDeGroot} as
\begin{equation}
	\frac{dN}{dy} = \lambda (1 - x_{+})^\lambda (1 - x_{-})^\lambda
					\label{eq:clandensity}
\end{equation}
with
\begin{equation}
	x_{\pm} \equiv \frac{m}{\sqrt{s}} e^{\pm y},
\end{equation}
where $m$ is the (average)
\textit{transverse} mass of the clan ($m_T = \sqrt{m^2 + p_T^2}$, it will be
called $m$ in following), $\lambda$ is a parameter
closely related to the plateau height, that is, 
to the average number of clans
per unit (pseudo)-rapidity $y$, and $\sqrt{s}$ is the c.m.\ energy.
Notice that Eq.~(\ref{eq:clandensity}) limits clan emission
to the interval $|y| < \ln(\sqrt{s}/m)$.

Assuming furthermore that each clan produces particles according to a
logarithmic MD, whose generating function is
\begin{equation}
	g_{\text{log}}(z) \equiv \sum_{n=1}^{\infty} z^n P_n^{\text{L}} (\beta)
			= \frac{\ln(1-z\beta)}{\ln(1-\beta)} ,
\end{equation}
with the average multiplicity per clan, $\nc$, given by
\begin{equation}
	\nc = \frac{\beta}{(\beta-1)\ln(1-\beta)} ,
\end{equation}
then the generating function for the MD in the (pseudo)-rapidity
interval $\Delta \eta$ turns out to be \cite{Bialas:clanmass} 
\begin{equation}
	G(z; \Delta \eta) = \exp\left\{
			\int \frac{dN}{dy} \frac{\ln\left[ 
					1 - \frac{\beta}{1-\beta} p(y;\Delta \eta) (z-1)
			\right]}{\ln(1-\beta)}  \;dy 
  \right\} ;
				\label{eq:1}
\end{equation}
$p(y;\Delta \eta)$ is the fraction of particles, produced by a
clan of (pseudo)-rapidity $y$, falling within $\Delta \eta$;
it was also assumed that for a fixed clan multiplicity the MD of particles
falling within $\Delta \eta$ is binomial, i.e., particles emitted
by each clan are emitted independently from each other. 
The integration is over the full range allowed by the kinematical limits
of Eq.~(\ref{eq:clandensity}).
The Authors of Ref.~\cite{Bialas:clanmass} assumed 
further, for the probability density function within a clan
to produce a particle at $\eta$, given that the clan is at $y$,
a form based on the hypothesis of isotropic decay:
\begin{equation}
	\phi(\eta; y) = \left[ 2 \omega \cosh^2\left( \frac{\eta-y}{\omega} \right) 
								\right]^{-1} 
																		\label{eq:2}
\end{equation}
($\omega = 1$ gives isotropic decay if $\eta$ is pseudo-rapidity;
the width of the distribution is proportional to $\omega$)
and thus computed
\begin{equation}
	p(y;\Delta \eta) = \int_{\Delta \eta} \phi(\eta; y) \;d\eta .
\end{equation}

There are, in summary, 4 free parameters to be used to fit the 
experimental data: $\beta$, $\lambda$, $\omega$ and $m$.

Experimental data on $p\bar p$ collisions at 546 GeV  
are approximately reproduced 
with the following  choice of the parameters in the generalised model
of Ref.~\cite{Bialas:clanmass}
\begin{equation} 
  \lambda= 0.855,\quad m= 3.15,\quad \omega= 1.45,\quad \beta=0.90.
\end{equation}

The obtained multiplicity distribution is not indeed of NB type 
except in full phase space
(deviations are significant for $n< 3$, and in $k$ 
parameter rapidity dependence 
at the border of phase space). A later analysis of these data 
\cite{frascati97:AG} has shown
indeed that  a shoulder effect occurs in the multiplicity distribution
and that one should try to explain observed data 
in  $n$-charged particle  multiplicity distributions in rapidity in terms
of the weighted superposition of two classes of events [soft (no mini-jets)
semi-hard (with mini-jets)] each described by a MD of NB (Pascal) type.
This consideration notwithstanding, the study in
Ref.~\cite{Bialas:clanmass} is, in our opinion,
very instructive because it allows to determine
characteristic clan parameters, like its width and mass, under 
simple  assumptions and it  can be used to determine important properties
of the  same   parameters in different classes of events and in
different collisions.

The answer to the first question is therefore positive.

\section{Second question.}

This result led us to the next question.
We studied first $pp$ collisions at 63 GeV in rapidity intervals:
according to our knowledge, at such c.m.\ energy only one component
(the soft one) is usually assumed to control the dynamics of the
collision and the shoulder effect is, to a good approximation, negligible.

\begin{figure}
  \begin{center}
  \mbox{\includegraphics[width=0.95\textwidth]{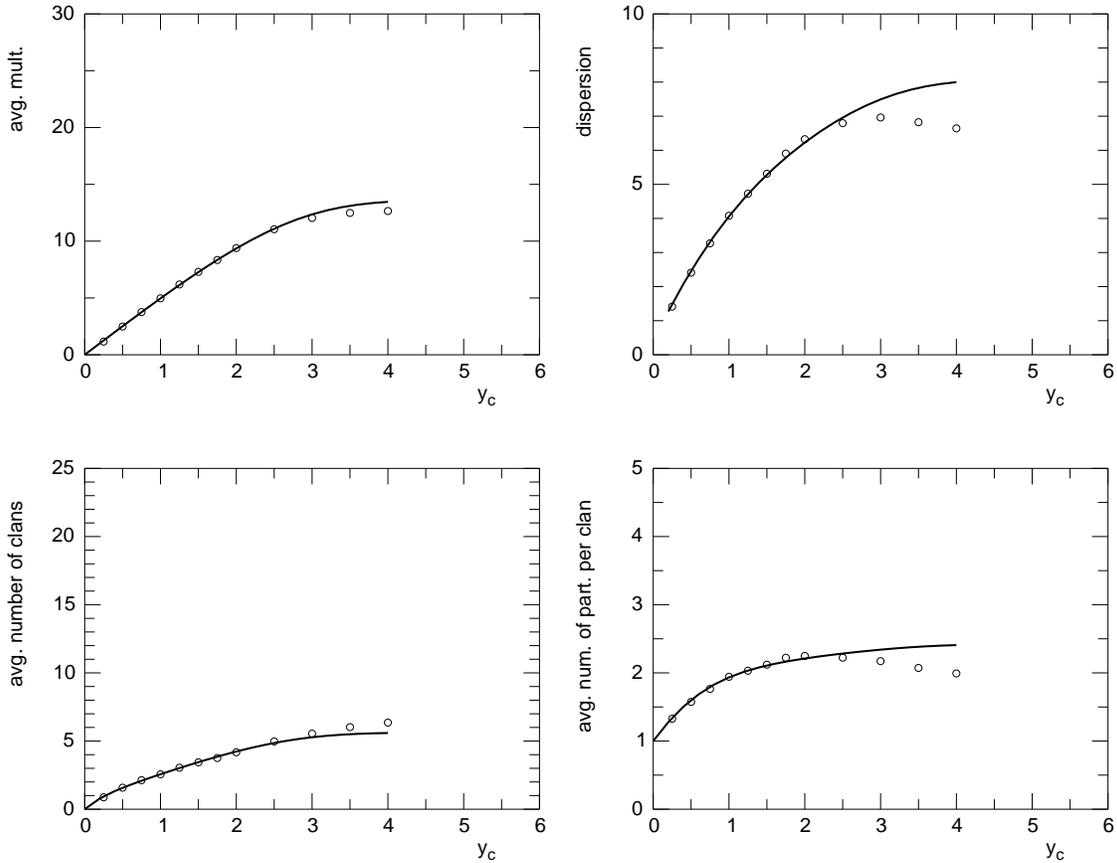}}
  \end{center}
  \caption{Average multiplicity, dispersion, average number of clans
  and average number of particles per clan
	in different pseudo-rapidity intervals $[-y_c,y_c]$ for the soft
  component of the MD in $pp$ collisions at 63 GeV: 
  open circles are  data \cite{ISR:0} and the solid line our
	fit.}\label{fig:fit63}
  \end{figure}

The average number of particles per clan, the average number of clans,
the dispersion, the parameter $k$ and charged particle multiplicities in
pseudo-rapidity intervals $\eta_c < 2.5$  are quite well reproduced
by the set of parameters shown in Table~\ref{tab:1}\textit{a}:
they are obtained by fitting, with the least
square method, the average multiplicity $\avg{n}$ and the
quantity $D^2/\avg{n}^2 - 1/\avg{n}$ of the distribution obtained
from the generating function (\ref{eq:1}), in terms of the four
parameters $\lambda$, $m$, $\omega$ and $\beta$,
to the corresponding moments of the NBMD (namely $\nbar$ and $k^{-1}$,
respectively)
for pseudo-rapidity intervals $\Delta\eta = [-\eta_c,\eta_c]$ with
$\eta_c \leq 2.5$ (see Fig.~\ref{fig:fit63}).

\begin{table}
  \caption{Parameters $\lambda$, $m$, $\omega$ and $\beta$,
  obtained from fitting Eq.~(\ref{eq:1}) to data 
	for each component at various
	c.m.\ energies and in different classes of collisions,
	are shown in the top part; other quantities
	derived from the fit parameters are show in the bottom part.}\label{tab:1}
	\begin{center}
  \begin{tabular}{l|c|cc|cc}
            \multicolumn{1}{c}{}
	          & \multicolumn{1}{c}{\textit{(a)}} & 
                              \multicolumn{2}{c}{\textit{(b)}} &
                              \multicolumn{2}{c}{\textit{(c)}}\\
            & $pp$ collisions
            & \multicolumn{2}{c|}{$p\bar p$ collisions} &
                        \multicolumn{2}{c}{\ee\ annihilation}\\
	          & 63 GeV & \multicolumn{2}{c|}{ 900 GeV} & 
	                      \multicolumn{2}{c}{ 91 GeV}\\
            & soft & ~~~soft~~~ & semi-hard & ~~~2-jet~~~ & 3-jet \\
  \hline						               		                 
	$\lambda$       & 1.14 & 0.92 & 2.09 	& 1.60 & 3.33 \\
	$m$ (GeV/$c^2$) & 1.80 & 1.47 & 3.43 	& 0.62 & 1.10 \\
  $\omega$        & 0.84 & 1.95 & 1.35 	& 1.34 & 0.56 \\
  $\beta$         & 0.79 & 0.83 & 0.87  & 0.62 & 0.59 \\
  \hline
	$dN/dy|_{y=0}$  & 1.15 & 0.92 & 2.06  & 1.57 & 3.07 \\
  $\Nbar$         & 5.59 & 9.83 & 16.8  & 11.0 & 16.6 \\
  $\nc$           & 2.41 & 2.69 & 3.21  & 1.64 & 1.62 \\
  \hline
  \end{tabular}
	\end{center}
  \end{table}

We remark that 

~\textit{a.} soft events only are considered in $pp$ at 63
GeV c.m.\ energy (in fact, one component only determines reasonably
well the $n$-charged particle multiplicity distribution in rapidity
intervals);

~\textit{b.} the average number of particles per clan,
$\nc$, is not correctly reproduced for rapidity intervals larger than
$\eta_c = 2.5$; a clear bending of $\nc$ is visible in the data and not
its increase with the rapidity interval considered as shown in the
generalised model (clans in central rapidity regions are larger than
in more peripheral regions at the border of phase space \cite{AGLVH:1}).

This fact has important consequences on the determination of the leakage
parameter introduced \cite{RU:FB} in the study 
of particles generated from clans lying in one hemisphere to
the opposite one  in forward-backward multiplicity correlations  and suggests
that the leakage  parameter  should be larger in broader clans than in smaller 
ones (and not the same throughout all the allowed rapidity range as done in 
\cite{RU:FB}).
The proposed value of the leakage parameter in the just mentioned 
reference should therefore be considered rather an average value  between
leakage parameters of large and small clans than  a rapidity independent  
value. 

\begin{figure}
  \begin{center}
  \mbox{\includegraphics[width=0.95\textwidth]{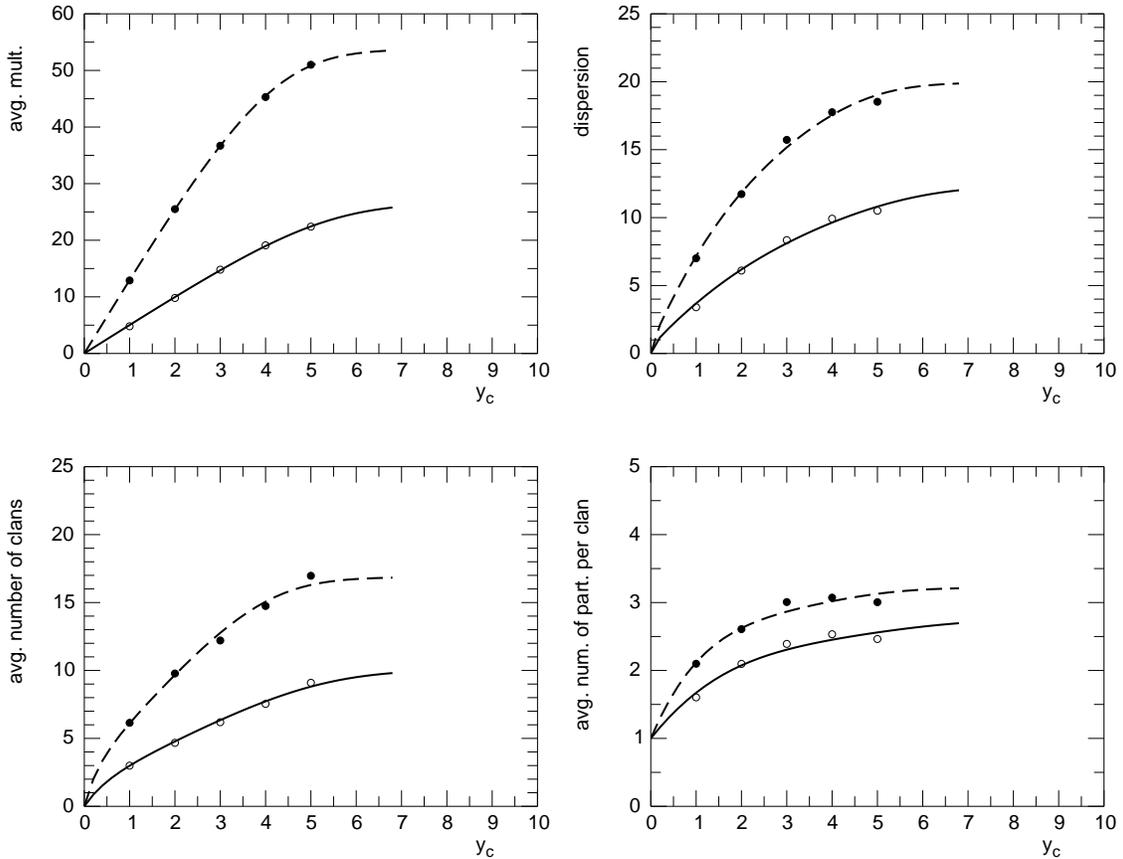}}
  \end{center}
  \caption{Fit to the average multiplicity and dispersion 
	in different pseudo-rapidity intervals $[-y_c,y_c]$ for the two
  components of the MD in $p\bar p$ collisions at 900 GeV:
	soft component: open circles (data) and solid line (fit); 
	semi-hard component: filled circles (data) and dashed 
  line (fit). Data points are from Ref.~\cite{Fug}.}\label{fig:fit}
  \end{figure}

\begin{figure}
  \begin{center}
  \mbox{\includegraphics[width=0.8\textwidth]{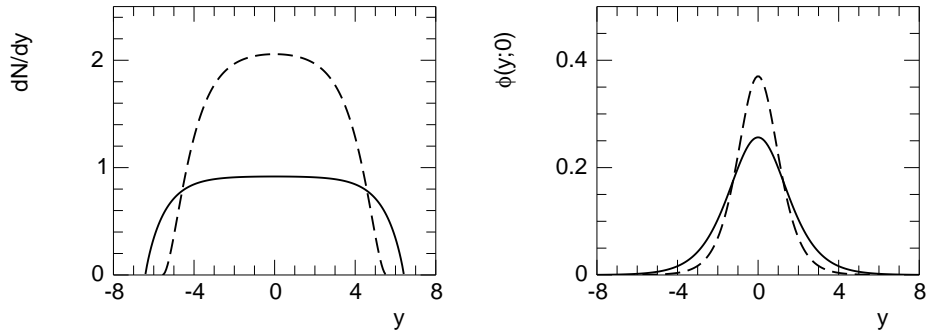}}
  \end{center}
  \caption{Clan density $dN/dy$, Eq.~(\ref{eq:clandensity}),
  and single particle 
	pseudo-rapidity probability density in a clan $\phi(y;0)$,
	Eq.~(\ref{eq:2}),
	for the soft (solid line) and semi-hard (dashed line)
	component at 900 GeV c.m.\ energy; parameters from 
	Table~\ref{tab:1}\textit{b}.}\label{fig:densities}
  \end{figure}

We decided then to study  $p\bar p$ collisions at 900 GeV c.m.\ energy.
According to our experience  here  not only one but  two components are
controlling the dynamics of the process, i.e.,  the soft and the semi-hard 
component \cite{combo:prd}.
We fit therefore the four parameters $\lambda$, $\omega$, $m$ and
$\beta$, with the method previously explained,
for each component separately using available UA5 data \cite{Fug},
i.e., the NB fits in pseudo-rapidity intervals $[-\eta_c,\eta_c]$
with $\eta_c = 1 \dots 5$;
because, as discussed in \cite{AGLVH:1},
clans emitted close to their kinematical limit $\ln(\sqrt{s}/m)$
appear to be smaller, we do not make use of full phase space
values.
The fits turn out to be good, as shown in Fig.~\ref{fig:fit},
except that they do not reproduce well the decrease in
the average number of particles per clan close 
to full phase space.
The resulting values of the fit parameters are shown 
in Table~\ref{tab:1}\textit{b}.
Figure~\ref{fig:densities} shows the densities 
$dN/dy$, Eq.~(\ref{eq:clandensity}),
and $\phi(\eta;0)$, Eq.~(\ref{eq:2}), for the two components
at 900 GeV separately, using the results of the fit.

% in the following
% table (first half) together with other variables derived from them (second
% half):
% \begin{center}
% \begin{tabular}{lcc}
%             & soft & semi-hard \\
%   \hline		              
% 	$\lambda$ & 0.92 & 2.09 \\
% 	$m$ (GeV) & 1.47 & 3.43 \\
%   $\omega$  & 1.95 & 1.35 \\
%   $\beta$   & 0.83 & 0.87 \\
%   \hline
% 	$dN/dy|_{y=0}$ & 0.92 & 2.06 \\
%   $\Nbar$        & 9.83 & 16.8 \\
%   $\nc$          & 2.69 & 3.21 \\
%   \hline
%   \end{tabular}
% \end{center}

In comparing parameters behaviour for the soft component at 63 GeV 
and  900 GeV we notice that  clan masses and  distribution
widths vary with c.m.\ energy and the plateau height is
slowly decreasing. The decrease of the clan mass in the soft component
with increasing energy could be due to the intentionally overlooked
contamination of semi-hard events at 63 GeV, as will be discussed in
the last section.

In comparing, next, parameters behaviour for the soft and semi-hard components
at the same  c.m.\ energy (900 GeV) we remark that clan masses and 
plateau heights  are much higher in the semi-hard than in the soft component,
whereas the distribution width is much higher in the soft than in the
semi-hard component. This fact shows  that heavier particles are
produced more in semi-hard than in the soft component.

Interestingly the average  number of particles per clan
is bending in larger rapidity intervals both  in the soft and
in the semi-hard component  suggesting also here  that clans are larger in 
central rapidity intervals than in the peripheral ones. 

Accordingly leakage parameters in forward backward multiplicity correlations 
should be larger  when clans have larger masses and their particle content is 
distributed  in more central rapidity intervals.  

In conclusion clan masses on the average are different in different
classes of events or substructures of a given class of collision.

\section{Third question.}

It is interesting to check clan properties also in \ee\ annihilation at
LEP c.m.\ energy  within the generalised model of Bialas and Szczerba.

We apply it to the 2-jet and 3-jet samples of events  and we find
that  data  are approximately reproduced with the choice 
of parameters shown in Table~\ref{tab:1}.

% \begin{center}
% \begin{tabular}{lcc}
%             & 2-jet & 3-jet \\
%   \hline		                
% 	$\lambda$ & 1.60 & 3.33 \\
% 	$m$ (GeV) & 0.62 & 1.10 \\
%   $\omega$  & 1.34 & 0.56 \\
%   $\beta$   & 0.62 & 0.59 \\
%   \hline
% 	$dN/dy|_{y=0}$ & 1.57 & 3.07 \\
%   $\Nbar$        & 11.0 & 16.6 \\
%   $\nc$          & 1.64 & 1.62 \\
%   \hline
%   \end{tabular}
% \end{center}

We conclude in comparing the average masses of 2- and 3-jet samples of
events in \ee\ annihilation at LEP energy with the average mass of the
soft component at 63 GeV in $hh$ collisions that clan masses in 
2-jet and 3-jet events are much lower than in the soft component at
ISR energies.

\section{Remarks and questions for future experiments}

Assuming the validity of NB (Pascal) regularity and its interpretation
in terms of clan structure for each substructure  (or component or
subsample of events) characterising different classes of high energy
collisions it was shown in previous works that the weighted
superposition  of NB (Pascal) MD's (one for each component) describes
quite well experimental final charged particle MD properties which a
single NB (Pascal) MD is unable to reproduce.
It should be pointed out that QCD has (up to now) quite poor
prediction on all just mentioned experimental facts:
experts claim that they are all consequences of the onset of hard
gluon radiation and rely fully on purely complicated higher order
perturbative calculations to be eventually performed in the future.
From a stricter, and complementary, phenomenological 
point of view, the above mentioned
successes in describing data in terms of the weighted superposition
mechanism of two (or eventually more) NB (Pascal) MD's and the related
interpretation in terms of clans of the different substructures of a
collision raise intriguing problems, which we summarised in the
following question: are clans observable objects?

This paper is an attempt to answer this interesting question.
It concerns mainly energy and rapidity dependence of average clan
masses of the substructures characterising \ee\ annihilation and $hh$
collisions. Results are not inconsistent with general expectations and
support the idea that clans could be indeed observable.

Some warnings are needed. Our work is based on informations and
extrapolations not coming from dedicated experiments on the subject,
as these are not available at present. We are therefore aware of the
intrinsic limitations of our approach.
For instance the separation between soft and semi-hard events at 900
GeV in $hh$ collisions comes (as already pointed out) from a fit
\cite{Fug}
proposed in order to reproduce observed experimental data on MD's.
The question still on the carpet
%MerriamWebster: on the carpet : before an authority for censure or reproof
is indeed what is a soft and what is a semi-hard event in $hh$
collisions.
Some progress has been made by CDF Collaboration
\cite{CDF:soft-hard}, but a definite generally
accepted answer is lacking.
Similarly at 63 GeV c.m.\ energy (ISR) in view of the lack of a clean
separation between soft and semi-hard events, all events were
taken to be soft (and consequently a single NB (Pascal) MD was used
for describing the full sample of events).
% but one NB works well!
It cannot be excluded that the relatively high average clan mass found
in the soft component at 63 GeV c.m.\ energy ($m \approx 1.80$ GeV/$c^2$) (no
semi-hard events were assumed to contribute to the full sample of
events) with respect to that of the average clan mass of the soft
component at 900 GeV ($m\approx 1.47$ GeV/$c^2$) (here both soft and semi-hard
events [according to Fuglesang's fit] were assumed to contribute to
the full sample of events) might be the consequence of the
contamination of a certain percentage of semi-hard events at 63 GeV
which at present we are unable to disentangle and which will modify
the calculated average clan mass of the soft component.

In addition, the separation between 2-jet and 3-jet events in \ee\
annihilation is jet-finder algorithm dependent and it might be that
one should consider the 3-jet sample events separated into two extra
subsamples of events (e.g., Mercedes-like events and hard-gluon
events).
All these remarks make of course our conclusions questionable.
These consideration notwithstanding, it seems to us interesting to
explore the possibility that clans do have mass.
This search should be done in our opinion in future experiments in two
steps.

Firstly, it is important to check with dedicated experimental work
existing substructures in the full sample of events of each class of
collisions and verify then that they agree with NB (Pascal) behaviour
or at least with infinitely divisible distribution (IDD) behaviour
(which maintains the Poissonian nature of the first step of the
production process).
Secondly, one should apply the analysis of the present approach to
each component in order to determine clan masses properties.
In this respect it is relevant to remember that the average masses we
are referring to are average transverse masses and therefore one should
pay attention to the transverse momentum component fraction
contributing to the real average clan mass which might be disturbing
in the separation between soft and semi-hard events.

It is clear that if clans are massive and their average masses vary
with energy and rapidity the next question is: do clans have other
other quantum  numbers?
If clans are real objects, do they continue to be independently
produced  and to decay with no correlations among particles generated
by different ancestors at all c.m.\ energies, or there exists a
threshold beyond which clans loose their original independence, start
to overlap and then to interact among themselves, modifying the
initial simplicity of clan structure interpretation of NB (Pascal), or
IDD, regularities?

Other interesting remarks and subtle questions can be added to the
previous ones.
They are too premature in our opinion. The point we want to make,
following the results of this approach, is that if clans are massive
objects (with eventually other quantum numbers) and their masses vary
with c.m.\ energy and rapidity, and clan masses are different in
different substructures of high energy collisions, the hadronization
process itself should be consistent with the new scenario which
emphasises the role of clans with respect to that of final charged
particles. 
The hadronization process in this perspective will be dominated by
massive clans (primaries?) whose search and characteristic properties
study will be a possible new frontier in multiparticle dynamics in all
high energy collisions, a challenging problem to experimentalists of
the next generation accelerators.

% \section*{Acknowledgements}

% \section*{References}
\bibliographystyle{prstyR}  % new, using bibref
\bibliography{abbrevs,bibliography}

\end{document}